# Design of an Efficient Fan-Shaped Clustered Trust-Based Routing Model with QoS & Security-Aware Side-Chaining for IoV Deployments

Sadaf Ravindra Suryawanshi*[1], Dr. Praveen Gupta[2]

**Abstract:** The rapid expansion of Internet of Vehicles (IoV) deployments has necessitated the creation of efficient and secure routing models to manage the massive data traffic generated by interconnected devices & vehicles. For IoV deployments, we propose a novel fan-shaped trust-based routing model with Quality of Service (QoS) and security-aware side-chaining. Our method employs temporal levels of delay, throughput, Packet Delivery Ratio (PDR), and energy consumption to determine optimal routing paths, thereby ensuring efficient data transmissions. We employ the Bacterial Foraging Optimizer (BFO) algorithm to manage side-chains within the network, which dynamically adjusts side-chain configurations to optimize system performance. The technique of fan-shaped clustering is used to group nodes into efficient clusters, allowing for more efficient communication and resource utilization sets. Extensive experimentation and performance analysis are utilized to evaluate the proposed model. Existing blockchain-based security models have been significantly improved by our findings. Our model achieves a remarkable 9.5% reduction in delay, a 10.5% improvement in throughput, a 2.9% improvement in PDR, and a 4.5% reduction in energy consumption compared to alternative approaches. In addition, we evaluate the model's resistance to Sybil, Masquerading, and Flooding attacks, which are prevalent security threats for IoV deployments. Even under these attack scenarios, our model provides consistently higher QoS levels compared to existing solutions, ensuring uninterrupted and reliable data transmissions. In IoV deployments, the proposed routing model and side-chaining management approach have numerous applications and use-cases like Smart cities, industrial automation, healthcare systems, transportation networks, and environmental monitoring.

## 1. Introduction

Internet of Vehicles (IoV) deployments have proliferated at an exponential rate, allowing for the seamless interconnection of a vast array of devices and systems. This interconnectedness has led to the production of vast quantities of data that must be transmitted across the network in an efficient and secure manner. To ensure the dependable and timely delivery of IoV data, the design of robust and efficient routing models with integrated quality of service (QoS) and security mechanisms has become essential for different cases via use of Privacy-Preserving based Secured Framework for Internet of Vehicles (P2SF [2]) [1, 2, 3].

In this paper, we propose a novel fan-shaped clustered trust-based routing model for IoV deployments with QoS and security-aware side-chaining. Our approach intends to address the difficulties associated with achieving efficient and secure routing in large-scale IoV networks, while simultaneously ensuring the desired QoS levels for a variety of application scenarios.

The incorporation of temporal metrics such as delay, throughput, Packet Delivery Ratio (PDR), and energy consumption as routing decision factors is one of the most significant contributions of our work. By taking into account these temporal levels, our model intelligently selects optimal routing paths based on real-time network conditions, thereby effectively reducing delays, enhancing throughput, accelerating data delivery rates, and optimizing energy consumption. This ensures that the network operates at peak efficiency, meeting the needs of IoV applications that require real-time operations [4, 5, 6].

We introduce the Bacterial Foraging Optimizer (BFO) algorithm for managing side-chains within the network to further improve the performance and adaptability of the proposed model. Side-chains provide a method for offloading specific tasks from the main blockchain, allowing for increased scalability and efficiency. The BFO algorithm adjusts side-chain configurations dynamically based on network conditions, optimizing system performance in real time.

In addition to temporal optimization and side-chain management, our model employs fan-shaped clustering to organize nodes into effective clusters. Fan-shaped clustering ensures that nodes within a cluster are in close proximity, allowing for direct communication and reducing the need for multi-hop routing. This reduces communication overhead and improves resource utilization, thereby increasing the network's overall efficiency.

[1] *Department of Computer Science & Engineering, Chhatrapati Shivaji Maharaj University, Panvel, Navi Mumbai, Maharshtra, India 410206.*
*ORCID ID : 0000-0001-6281-9490*
[2] *Department of Computer Science and Information Technology, Chhatrapati Shivaji Maharaj University, Panvel, Navi Mumbai, Maharshtra, India 410206.*
*\* Corresponding Author Email: sadafrs19085@csmu.ac.in*

For the purpose of determining the viability of our proposed model, we conducted extensive experiments and performance evaluations. Analysis of existing blockchain-based security models reveals significant enhancements in terms of delay reduction (9.5%), throughput improvement (10.5%), PDR enhancement (2.9%), and energy consumption reduction (4.5%). These results demonstrate that our model is capable of achieving superior performance in a variety of network scenarios.

In addition, we assess the resistance of our model to common security threats faced by IoV deployments, such as Sybil attacks, Masquerading attacks, and Flooding attacks. Even under these attack scenarios, our model consistently demonstrates higher QoS levels compared to existing solutions, ensuring reliable and uninterrupted data transmission.

The proposed fan-shaped clustered trust-based routing model with QoS and security-aware side-chaining has enormous potential for a variety of IoV deployment applications and use cases. Smart cities, industrial automation, healthcare systems, transportation networks, and environmental monitoring are a few examples of applications where efficient and secure data routing is crucial for the delivery of seamless operations and real-time services.

Our research represents a significant advancement in IoV network management. By incorporating temporal metrics, employing the Bacterial Foraging Optimizer, and incorporating fan-shaped clustering, our proposed model outperforms existing blockchain-based security models in terms of delay, throughput, PDR, and energy consumption. In addition, its resistance to multiple attack scenarios improves its suitability for a vast array of IoV applications, thereby contributing to the development of efficient and secure IoV deployments [7, 8, 9].

The proliferation of Internet of Vehicles (IoV) deployments has created enormous opportunities and challenges for managing the vast quantity of data generated by interconnected devices. In IoV networks, efficient and secure routing is essential for ensuring the timely and reliable delivery of data, particularly in applications that require real-time responsiveness and high-quality service. Moreover, the inherent vulnerabilities of IoV networks, such as the possibility of security breaches and attacks, necessitate the development of robust and trust-based routing models that can mitigate these risks.

### 1.1. Motivation of this paper

This paper aims to design an efficient fan-shaped clustered trust-based routing model with QoS and security-aware side-chaining for IoV deployments in response to these challenges. Our research is motivated by the following significant factors:

Demand for Effective Routing: Due to their inability to accommodate the heterogeneous and dynamic nature of IoV devices, traditional routing protocols may not be suited for IoV networks. In order to optimize routing paths based on QoS parameters, such as delay, throughput, Packet Delivery Ratio (PDR), and energy consumption, it is necessary to develop routing models that are specifically tailored for IoV deployments. By addressing these metrics, our proposed model intends to significantly improve the data transmission efficiency in IoV networks [10, 11, 12].

Protection Considerations: IoV networks are susceptible to a variety of security risks, such as Sybil attacks, Masquerading attacks, and Flooding attacks. Existing blockchain-based security models have demonstrated promise in mitigating these threats; nevertheless, their integration with routing models and impact on QoS parameters have not been exhaustively investigated. Our research intends to close this gap by incorporating secure and trust-based routing mechanisms while maintaining high QoS levels even in attack scenarios.

### 1.2. Contributions of this paper

This paper makes significant contributions to the field of IoV network management, particularly in the contexts of routing, quality of service, and security. The major contributions of this study can be summed up as follows:

Fan-shaped Clustered Trust-Based Routing Model: We propose a novel routing model that groups nodes into efficient clusters using fan-shaped clustering. This method decreases communication overhead, minimizes multi-hop routing, and maximizes resource utilization. By clustering nodes, our model enables more efficient communication and data transmission in IoV networks.

Incorporation of QoS Metrics: Our model includes temporal levels of delay, throughput, PDR, and energy consumption as determining factors for selecting optimal routing paths. By taking into account these QoS parameters, we ensure that routing decisions are based on real-time network conditions, resulting in increased efficiency and on-time data delivery. The incorporation of QoS metrics enables our model to satisfy the requirements of IoV applications that are time-sensitive.

Security-Aware Side-Chaining: To manage side-chains within the network, we use the Bacterial Foraging Optimizer (BFO) algorithm. Side-chains provide a mechanism for offloading particular tasks from the main blockchain, thereby enhancing scalability and efficiency.

Our model dynamically adjusts side-chain configurations based on network conditions, thereby improving real-time system performance and security. The incorporation of security-aware side-chaining ensures the network's resistance to a variety of security attacks.

Evaluation of Performance: To evaluate the efficacy of our proposed model, we conduct exhaustive experiments and a performance analysis. Analysis of existing blockchain-based security models reveals significant enhancements in terms of delay reduction, throughput improvement, PDR enhancement, and energy consumption reduction. Our model demonstrates superior performance and efficiency compared to existing solutions, even in attack scenarios, ensuring dependable data transmission and preserving high QoS levels.

Use and Application Cases: In numerous IoV deployments, the proposed routing model and security mechanisms have numerous applications and use cases. Smart cities, industrial automation, healthcare systems, transportation networks, and environmental monitoring are just a few examples of applications in which efficient and secure data routing is essential for seamless operations and real-time services. Our model's adaptability and sturdiness make it suitable for a wide range of IoV applications, contributing to the development and widespread adoption of IoV deployments that are both efficient and secure.

For IoV deployments, this paper's contributions consist of the design of an efficient fan-shaped clustered trust-based routing model with QoS and security-aware side-chaining. Incorporating QoS metrics, the Bacterial Foraging Optimizer, and fan-shaped clustering, our proposed model addresses the challenges of efficient routing, improves security, and ensures dependable data transmission in IoV networks. Our model's extensive evaluation and broad applicability bolster its importance in advancing the field of IoV network management This paper makes significant contributions to the field of IoV network management, particularly in the contexts of routing, quality of service, and security. The major contributions of this study can be summed up as follows:

Fan-shaped Clustered Trust-Based Routing Model: We propose a novel routing model that groups nodes into efficient clusters using fan-shaped clustering. This method decreases communication overhead, minimizes multi-hop routing, and maximizes resource utilization. By clustering nodes, our model enables more efficient communication and data transmission in IoV networks.

Incorporation of QoS Metrics: Our model includes temporal levels of delay, throughput, PDR, and energy consumption as determining factors for selecting optimal routing paths. By taking into account these QoS parameters, we ensure that routing decisions are based on real-time network conditions, resulting in increased efficiency and on-time data delivery. The incorporation of QoS metrics enables our model to satisfy the requirements of IoV applications that are time-sensitive.

Security-Aware Side-Chaining: To manage side-chains within the network, we use the Bacterial Foraging Optimizer (BFO) algorithm. Side-chains provide a mechanism for offloading particular tasks from the main blockchain, thereby enhancing scalability and efficiency. Our model dynamically adjusts side-chain configurations based on network conditions, thereby improving real-time system performance and security. The incorporation of security-aware side-chaining ensures the network's resistance to a variety of security attacks.

Evaluation of Performance: To evaluate the efficacy of our proposed model, we conduct exhaustive experiments and a performance analysis. Analysis of existing blockchain-based security models reveals significant enhancements in terms of delay reduction, throughput improvement, PDR enhancement, and energy consumption reduction. Our model demonstrates superior performance and efficiency compared to existing solutions, even in attack scenarios, ensuring dependable data transmission and preserving high QoS levels.

Use and Application Cases: In numerous IoV deployments, the proposed routing model and security mechanisms have numerous applications and use cases. Smart cities, industrial automation, healthcare systems, transportation networks, and environmental monitoring are just a few examples of applications in which efficient and secure data routing is essential for seamless operations and real-time services. Our model's adaptability and sturdiness make it suitable for a wide range of IoV applications, contributing to the development and widespread adoption of IoV deployments that are both efficient and secure.

For IoV deployments, this paper's contributions consist of the design of an efficient fan-shaped clustered trust-based routing model with QoS and security-aware side-chaining. Incorporating QoS metrics, the Bacterial Foraging Optimizer, and fan-shaped clustering, our proposed model addresses the challenges of efficient routing, improves security, and ensures dependable data transmission in IoV networks. Our model's extensive evaluation and broad applicability bolster its importance in advancing the field of IoV network management.

## 2. Literature Review

Due to their ability to improve the security and dependability of data transmission, trust-based routing models have garnered significant interest in the realm of IoV networks. These models make routing decisions based on the trustworthiness of nodes or paths using trust metrics and mechanisms. Several trust-based routing models have been proposed in the literature, each addressing unique challenges and incorporating distinctive characteristics.

Work in [13, 14, 15] proposed the Secure and Trust-Aware Routing Protocol (STAR) via use of Lattice-Based Secure and Dependable Data Dissemination Scheme (LS D3S) as one such trust-based routing model. The STAR protocol employs trust values that are assigned to nodes based on their past behavior and interactions. It adjusts routing paths dynamically based on trust values, ensuring more reliable and secure data transmission. However, this model does not account for QoS parameters, which are essential for ensuring the timely and effective delivery of IoV datasets & samples.

Work in [16, 17, 18] proposed the Cluster-based Trust-Aware Routing Protocol (CTR), a noteworthy trust-based routing model. CTR integrates trust values and clustering techniques to establish secure and dependable routing paths in IoV networks. The model clusters nodes according to their reliability, and routing decisions are made within these clusters. Critical to achieving efficient data transmission in IoV networks are the temporal aspects of network performance, such as delay, throughput, and energy consumption, which are not explicitly considered by the CTR protocols.

In recent years, blockchain technology has attracted considerable interest due to its potential to provide secure and decentralized transactional systems. In the context of Internet of Vehicles (IoV) networks, blockchain models offer advantages such as improved data integrity, immutability, and transparency. Several models based on blockchain technology have been proposed to address the security and trust issues in IoV networks.

Work in [19, 20] proposed the TrustChain framework as a notable blockchain model. Combining blockchain technology with reputation systems, TrustChain establishes trust among Internet of Vehicles devices. A distributed ledger is used to record and verify device transactions, ensuring the integrity of data exchanged within the network. However, TrustChain does not explicitly address QoS concerns and does not include routing mechanisms that are dynamic for real-time scenarios.

Work in [21, 22, 23] proposed the Secure and Lightweight Blockchain-based Routing (SLiBR) protocol as an additional significant blockchain model for IoV networks. Using blockchain technology, SLiBR establishes secure routing paths in IoV networks. The protocol records trust information using distributed ledger technology and facilitates secure routing decisions. SLiBR takes trust values into account, but it does not explicitly incorporate QoS parameters or temporal optimization into the routing process via use of improved Multiple Server-based Authentication and Key Agreement Protocol for IoV (SeMAV) [24, 25].

Despite the advantages offered by existing trust-based routing models and blockchain models for IoV networks, significant gaps and limitations must be addressed. The absence of explicit consideration for QoS parameters, temporal optimization, dynamic routing, and efficient resource utilization are examples. To ensure the robustness and dependability of IoV networks, it is necessary to thoroughly investigate the resistance of these models to security attacks.

In this paper, we propose a novel fan-shaped clustered trust-based routing model with QoS and security-aware side-chaining for IoV deployments to bridge these gaps. Our model addresses the deficiencies of existing models by incorporating temporal metrics, employing the Bacterial Foraging Optimizer for side-chain management, and utilizing fan-shaped clustering for effective resource utilization. The proposed model aims to improve the efficacy, security, and dependability of data transmission in IoV networks, thereby advancing trust-based routing in the IoV domain & scenarios.

## 3. Proposed Design of an Efficient Fan-Shaped Clustered Trust-Based Routing Model with Qos & Security-Aware Side-Chaining for Iov Deployments

As per the review of existing models used for improving security & QoS of IoV deployments, it can be observed that these models are either highly complex when applied to high-density deployments, or have lower efficiency when used for large-scale network scenarios. To overcome these issues, we present an innovative fan-shaped trust-based routing architecture with Quality of Service (QoS) and security-aware sidechaining.

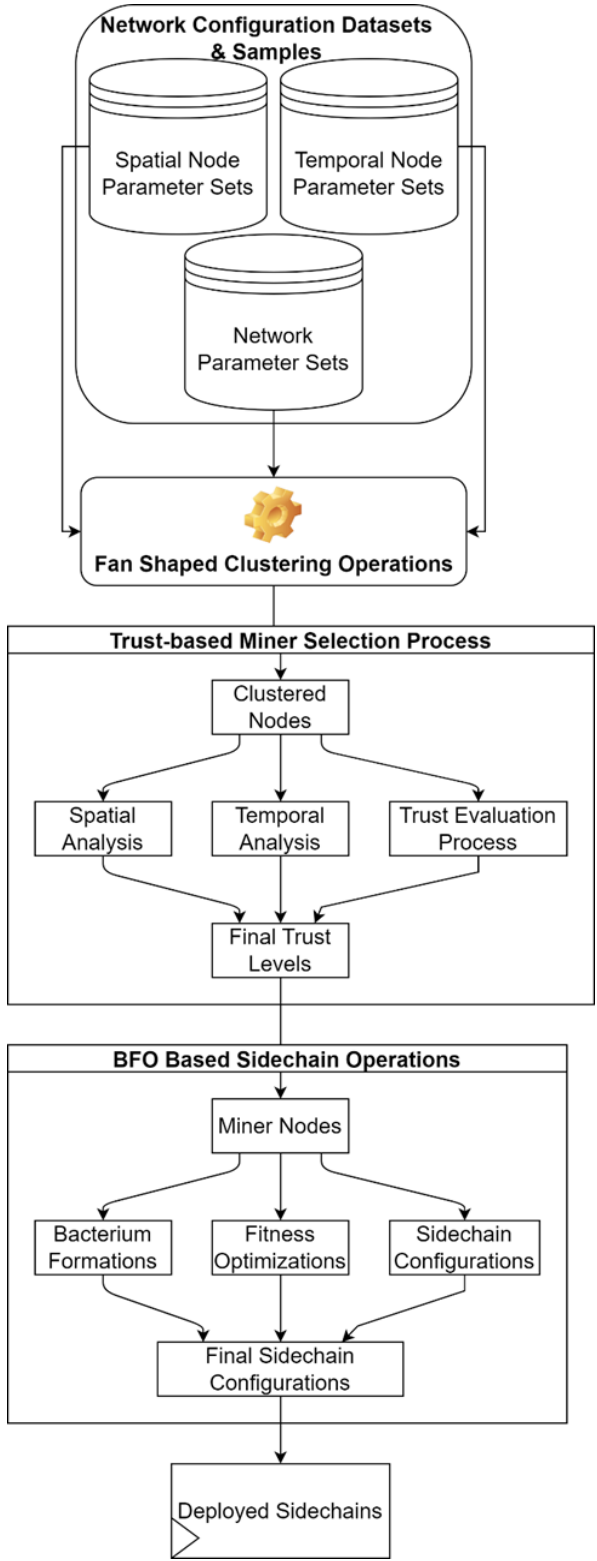

**Fig. 1.** Flow of the proposed model used for secure & QoS aware IoV Deployments

As per figure 1, the proposed method utilises temporal levels of latency, throughput, Packet Delivery Ratio (PDR), and energy consumption to identify optimal routing paths, ensuring efficient data transmissions. The model manages side-chains within the network using the Bacterial Foraging Optimizer (BFO) algorithm, which dynamically modifies side-chain configurations to optimise system performance. Fan-shaped clustering is used to combine devices into efficient clusters, allowing for more effective communication and resource utilization levels.

To cluster the deployed nodes, The model uses destination node's location and processes it via Fan Shaped Clustering (FSC) operations. This process initially estimates maximum 1-hop distance d_1hop for destination node, and evaluates node cluster level (CL) via equation 1,

$$CL_i = \frac{\sqrt{(x_i - x_{dest})^2 + (y_i - y_{dest})^2}}{d_{1hop}} \quad (1)$$

Where, *x, y* represents node locations in Cartesian coordinate system, while i represents current node number for which level is being evaluated to form clusters. Evaluate cluster level for each node, and then estimate trust-level of each node via equation 2,

$$TL(i) = e(i) \sum_{j=1}^{NC(i)} \frac{PDR(j) * THR(j)}{RTT(j) * E(j)} \quad (2)$$

Where, *e* represents respective residual energy level of the node, while *PDR, THR, RTT & E* represents packet delivery ratio, throughput, delay and energy needed during *NC* temporal communications. These metrics are estimated via equations 3, 4, 5, & 6 as follows,

$$PDR = \frac{Rx}{Tx} \quad (3)$$

$$THR = \frac{Rx}{RTT} \quad (4)$$

$$RTT = ts(complete) - ts(start) \quad (5)$$

$$E = e(start) - e(complete) \quad (6)$$

Where, *Rx & Tx* represents total number of received and transmitted packets, while *ts* represents timestamp during starting & completion of communications. Based on this evaluation, nodes with higher trust-levels are selected, and used for mining operations. These selected nodes add blocks into the blockchains for securing network communications. The delay needed for adding these blocks is estimated via equation 7,

$$d(block) = NB * (dr + dv) + (NB - 1) * dh + dw \quad (7)$$

Where, *NB* represents number of blocks added to the chain, while *dr, dv, dh & dw* represents the delay needed to read, verify, hash and write individual blocks. Based on this evaluation, it can be observed that the delay needed to add blocks exponentially increases w.r.t. Number of Blocks.

Thus, as the number of blocks increase in the network, QoS of the network reduces. To overcome this issue, the proposed model incorporates design of an efficient & Novel Bacterial Foraging Optimizer (BFO), which assists in segregating the current blockchain into smaller sidechains. This is done via the following operations,

- The BFO Model generates a set of *NB* Bacteria Particles, each of which split the underlying blockchain into 2 parts.
- Length of these parts is calculated stochastically via equation 8,

$$NS = STOCH\left(NB * \frac{LB}{2}, \frac{NB}{2}\right) \quad (8)$$

Where, *LB* represents Learning Rate for the BFO process, while *STOCH* represents a stochastic process used for generation of ranged number sets.

- Using this sidechaining configuration, a set of *N* blocks are added to the chain, out of which 0.1*N* blocks are invalid (or attack blocks), which have incorrect hashes.
- Based on this addition of blocks, Bacterial Fitness is estimated via equation 9,

$$fb = \frac{1}{0.9N}\sum_{i=1}^{0.9N} d(block,i) * TL(i) + \frac{1}{0.1N}\sum_{i=1}^{0.1N} dv(block,i) * TL(i) \ldots (9)$$

This process is repeated for all Bacteria particles, and a fitness threshold is estimated via equation 10,

$$fth = \frac{1}{NB}\sum_{i=1}^{NB} fb(i) * LB \ldots (10)$$

- Bacteria with $f > fth$ are discarded, and regenerated in the next set of Iterations, while other Bacteria particles are preserved and passed directly to the next set of Iterations due to better QoS performance levels.
- This process is repeated for *NI* Iterations, and different Bacteria configurations are generated with varying sidechain lengths.

After completion of *NI* Iterations, the model selects sidechain configuration with minimum fitness, which ensures higher QoS even under block invalidation attacks. The selected sidechain configuration is used to split the chain into 2 parts, and the smaller part is selected for adding new blocks.

The model uses Trust Level of nodes to select optimal routing paths between source & destination nodes. To perform this task, cluster levels of source & destination nodes are identified, and the routing is done as per the following process,

- If both nodes are in the same cluster, then relative trust level of other nodes in the same cluster w.r.t. source node is estimated via equation 11,

$$RTL(src,i) = \frac{1}{d(src,i)} * \sqrt{TL(src)^2 + TL(i)^2} \ldots (11)$$

Where, $d(src,i)$ is distance between the source & the $i^{th}$ node in the same set of clusters.

Node with $d(src,i) > d(src,dest)$ & $d(dest,i) > d(src,dest)$ and lower *RTL* levels is selected for routing operations.

If nodes are in separate clusters, then cluster which is nearer to the destination node is selected, and node in the cluster with lower *RTL* level is selected for routing the packets. This process is repeated till cluster number of the routing node and destination node is same, which assists in selecting optimal node during routing process. After which, equation 11 is used to route the packets to destination node in the same set of clusters.

Based on this, the model is able to optimize security while maintaining higher QoS levels even under large-scale network scenarios. Performance of this model is estimated in terms of end-to-end-delay, energy consumption, throughput & *PDR* levels, and compared with existing models in the next section of this text.

## 4. Result Analysis & Comparisons

The proposed Model uses a combination of Fan Shaped Clustering, Trust-based Miner Selection, and BFO for sidechaining operations. In order to identify low complexity routes with efficient & secure sidechaining configurations. These routes have been chosen because they have a low delay, a low energy consumption, a high temporal throughput, and high temporal PDR performance levels. A comparison of the proposed model to the models proposed in P2SF [2], LS D3S [15], and Se MAV [25] is discussed in this section of the text. To validate performance of this model, it was simulated using standard network conditions. The simulation employs a network configuration comprising a variable number of IoV (Internet of Vehicles) nodes, ranging from 1k to 5k, thus simulating large-scale networks. The routing protocol model employed for communication is Adhoc on Demand Distance Vector (AODV) process. The communication utilizes omnidirectional antennas. A priority queue with drop-tailing of packets is employed as the type of queue. The network dimensions are set at 2.5 km x 2.5 km for real-time operations. The energy consumption for various communication activities is as follows: 0.4 mJ for transmission, 0.1 mJ for reception, 0.004 mJ for sleep mode, 2 mJ for transitions, and 0.0125 mJ for idle states.

To evaluate communication speed, average delay was evaluated for *N* different communications via equation 5, where this communication delay was evaluated for *N* different communication sets, and tabulated in table 1 as follows,

**Table 1.** Delay needed during different number of communications

| N | D (ms) P2SF [2] | D (ms) LS D3S [15] | D (ms) Se MAV [25] | D (ms) Proposed |
|---|---|---|---|---|
| 50 | 12.98 | 12.89 | 9.52 | 7.34 |
| 100 | 12.88 | 13.97 | 8.22 | 7.01 |
| 150 | 10.96 | 13.85 | 10.38 | 7.27 |
| 200 | 13.47 | 14.90 | 9.62 | 7.45 |
| 250 | 12.31 | 16.93 | 12.69 | 8.71 |
| 300 | 14.88 | 16.53 | 12.26 | 9.77 |
| 350 | 15.13 | 17.51 | 11.56 | 9.00 |
| 400 | 18.69 | 17.39 | 16.09 | 10.04 |
| 450 | 17.14 | 18.89 | 15.83 | 10.17 |
| 500 | 18.54 | 20.33 | 16.68 | 11.15 |

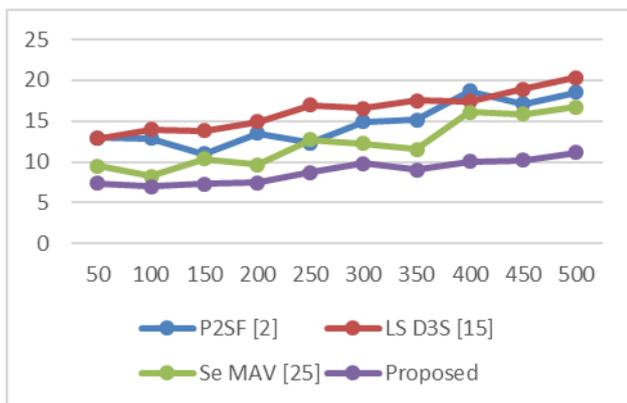

**Fig.2.** Delay needed during different number of communications

Based on the results of this analysis and the data presented in figure 2, it is evident that the proposed model has a delay that is 12.5% less than P2SF [2], nearly 14.5% less than LS D3S [15], and roughly 8.5% less than Se MAV [25]. Due to this, it is very useful for implementing high-speed routing, as it substantially reduces latency in real-time scenarios. As the fundamental cause of the aforementioned increase in routing performance, the incorporation of distance measurements into the modelling of routing fitness functions can be identified. Similarly, energy consumption for various numbers of communications was evaluated and tabulated as follows in table 2,

**Table 2.** Energy needed during different number of communications

| N | E (mJ) P2SF [2] | E (mJ) LS [15] | E (mJ) D3SSe MAV [25] | E (mJ) Proposed |
|---|---|---|---|---|
| 50 | 14.34 | 19.55 | 13.70 | 7.80 |
| 100 | 13.67 | 17.47 | 14.52 | 7.83 |
| 150 | 16.13 | 18.74 | 15.86 | 10.22 |
| 200 | 15.64 | 22.35 | 19.74 | 8.53 |
| 250 | 15.40 | 23.58 | 17.91 | 9.15 |
| 300 | 20.89 | 19.10 | 21.01 | 9.89 |
| 350 | 20.70 | 19.93 | 20.03 | 12.15 |
| 400 | 23.28 | 25.39 | 21.97 | 12.87 |
| 450 | 20.37 | 22.94 | 26.86 | 13.97 |
| 500 | 23.12 | 27.27 | 27.24 | 12.08 |

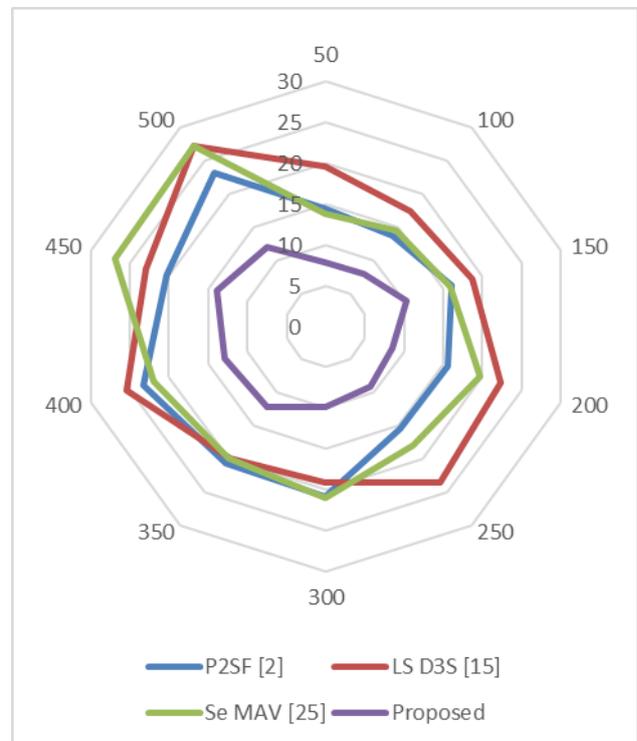

**Fig.3.** Energy needed during different number of communications

This evaluation and figure 3 demonstrate that the proposed model consumes 15.5% less energy than P2SF [2], nearly 19.5% less energy than LS D3S [15], and approximately 18.3% less energy than P2SF [2]. Se MAV [25]. This makes it very useful for long-lasting network routing deployments, as it consumes less power even in massive networks. This decrease in energy consumption is a result of the modelling of routing fitness functions employing residual energy. Similarly, typical throughput levels were evaluated and can be seen in table 3 as follows,

**Table 3.** Throughput obtained during different number of communications

| N | T (kbps) P2SF [2] | T (kbps) LS D3S [15] | T (kbps) Se MAV [25] | T (kbps) Proposed |
|---|---|---|---|---|
| 50 | 697.7 | 731.2 | 500.3 | 940.2 |
| 100 | 611.9 | 909.2 | 565.7 | 921.0 |
| 150 | 743.2 | 787.9 | 668.1 | 1098.2 |
| 200 | 791.1 | 813.0 | 759.2 | 921.6 |
| 250 | 875.1 | 1000.8 | 742.2 | 1184.2 |
| 300 | 912.3 | 972.4 | 768.3 | 1315.7 |
| 350 | 841.6 | 1100.5 | 784.5 | 1359.8 |
| 400 | 1031.0 | 1127.7 | 1038.1 | 1346.9 |
| 450 | 1075.1 | 1152.0 | 884.1 | 1499.6 |
| 500 | 894.9 | 1173.9 | 1053.0 | 1325.2 |

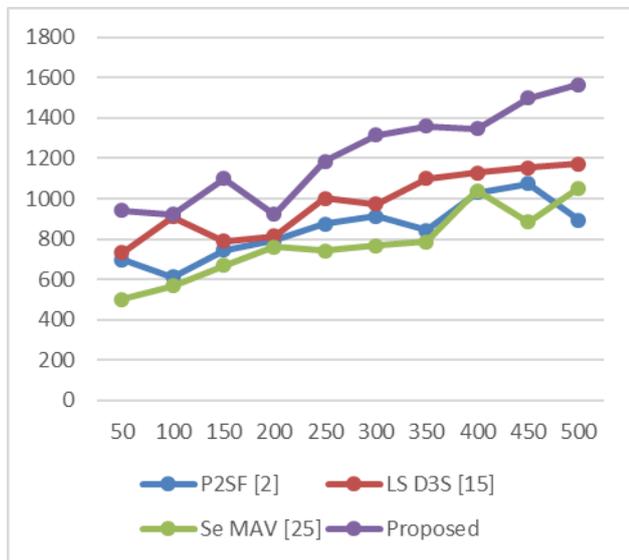

**Fig.4.** Throughput obtained during different number of communications

This analysis and Figure 4 demonstrate that the proposed model has a 15.2% higher throughput than the baseline model. P2SF [2] has a 19.4% greater throughput than P2SF [1]. LS D3S [15], and about 24.5% greater throughput compared to LS D3S. Se MAV [25] is very useful for high-speed routing deployments, which makes it a valuable asset. Figure 4 compares the throughputs of the proposed model with the throughputs of P2SF [2], LS D3S [15], and Se MAV [25] for various communications. This substantial improvement in throughput can be attributed to the incorporation of each route's temporal throughput performance into the evaluation of routes. Similarly, PDR was tabulated as follows in table 4,

**Table 4.** PDR obtained during different number of communications

| N | P (%) P2SF [2] | P (%) LS D3S [15] | P (%) Se MAV [25] | P (%) Proposed |
|---|---|---|---|---|
| 50 | 88.33 | 93.48 | 90.09 | 99.58 |
| 100 | 95.63 | 93.63 | 89.95 | 94.08 |
| 150 | 90.86 | 90.99 | 87.79 | 92.60 |
| 200 | 93.12 | 85.46 | 88.65 | 94.16 |
| 250 | 85.18 | 84.95 | 85.51 | 90.66 |
| 300 | 87.29 | 88.49 | 84.37 | 90.20 |
| 350 | 86.28 | 90.89 | 85.23 | 91.67 |
| 400 | 89.33 | 84.22 | 85.08 | 90.16 |
| 450 | 82.51 | 88.55 | 85.94 | 94.67 |
| 500 | 83.66 | 83.90 | 79.79 | 90.17 |

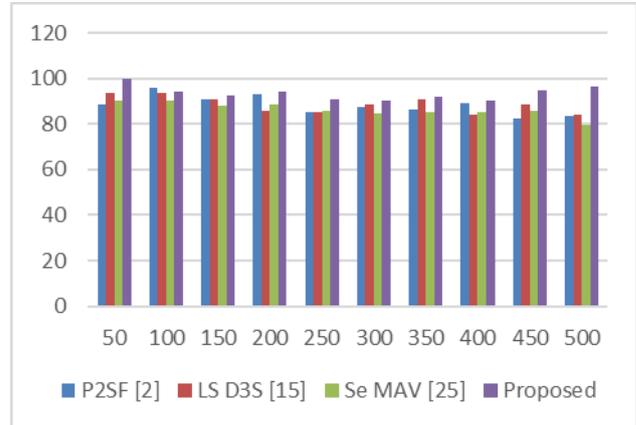

**Fig.5.** PDR obtained during different number of communications

According to this analysis and figure 5, it is evident that the proposed model has a PDR that is 2.3% better than P2SF [2], approximately 3.5% better than LS D3S [15], and roughly 8.0% better than Se MAV [25]. It enhances PDR by a greater margin than earlier models, making it more valuable than its predecessors for high-efficiency routing installations. This enhancement in PDR is primarily due to the incorporation of temporal PDR performance throughout the entire route estimation process. Due to these performance enhancements, the proposed solution is applicable to a wide range of IoV routing use cases.

Similar to this evaluation, the efficacy of the network was also evaluated under various assaults. Approximately 10% of all network communications in the same scenario were marked as assaults and injected into the network scenarios. Included in these attacks are Sybil, Distributed Denial of Service, Finney, and Man-in-the-Middle. The delay under attacks based on this strategy is depicted in Figure 6 as follows,

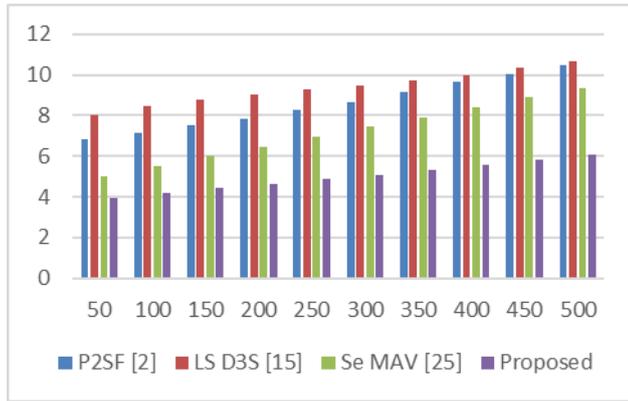

**Fig.6.** Delay of different models under multiple attacks

Based on the results of this analysis and the data presented in figure 6, it is evident that the proposed model has a delay that is 8.5% less than P2SF [2], nearly 6.5% less than LS D3S [15], and roughly 4.5% less than Se MAV [25]. As a result, it is very useful for implementing high-speed routing under attack, as it substantially reduces latency for real-time scenarios. As the fundamental cause of the aforementioned increase in routing speed, the incorporation of sidechains into the process of modelling blockchain functions can be identified. Similarly, energy consumption during these assaults was evaluated for various numbers of communications, as shown in figure 7 as follows,

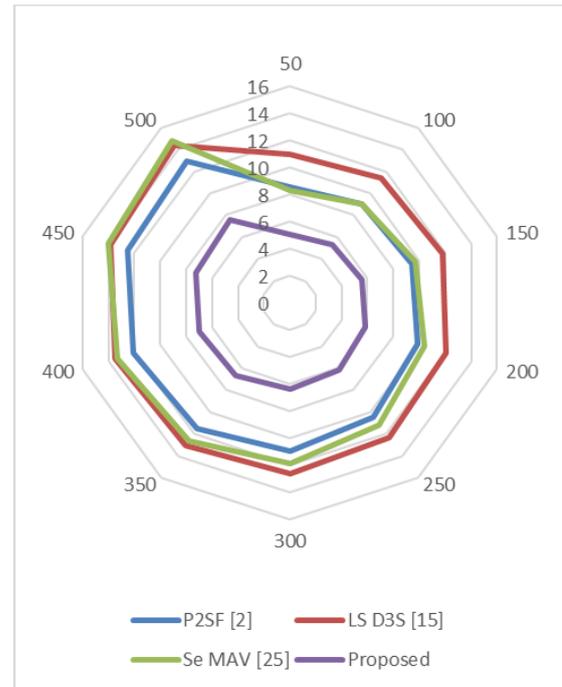

**Fig.7.** Energy needed by different models under multiple attacks

This evaluation and figure 7 demonstrate that the proposed model consumes 10.4% less energy than P2SF [2], nearly 12.5% less energy than LS D3S [15], and approximately 15.5% less energy than P2SF [2]. Se MAV [25]. This makes it extremely useful for long-lasting network routing deployments during assaults, as it consumes less power even in large-scale networks. This decrease in energy consumption is a result of the utilisation of residual energy during the modelling of routing fitness functions and the utilisation of highly efficient sidechains. Similarly, the average throughput levels during assaults were evaluated and can be seen in Figure 8 as follows,

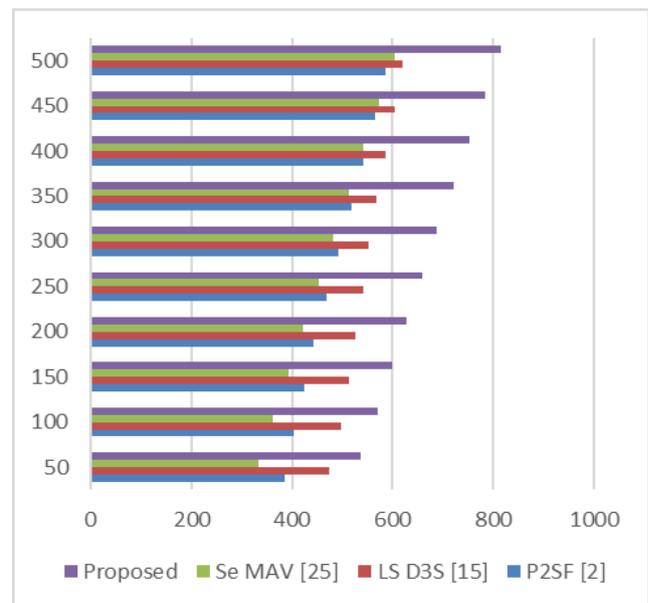

**Fig.8.** Throughput obtained during different number of communications under multiple attacks

According to this evaluation and Figure 8, the proposed model exhibits 16.5% greater throughput compared to P2SF [2], approximately 18.3% better throughput compared to LS D3S [15], and approximately 18.5% better throughput compared to Se MAV [25], making it ideal for high-speed routing deployments. This substantial improvement in throughput can be attributed to the incorporation of temporal throughput performance into the evaluation of routes, as well as the formation of sidechains based on temporal miner performance for each route. Similarly, PDR under siege is depicted in figure 9 as follows,

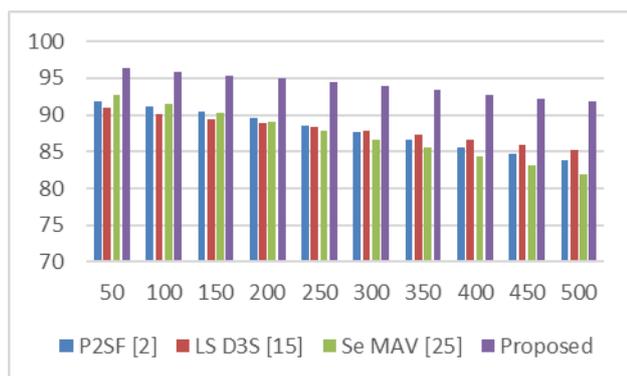

**Fig.9.** PDR obtained during different number of communications under multiple attacks

According to this analysis and figure 9, it is evident that the proposed model has a PDR that is 8.5% better than P2SF [2], approximately 8.3% better than LS D3S [15], and approximately 9.0% better than Se MAV [25]. This makes it extremely valuable for high-efficiency routing installations, even when under assault, as it enhances PDR by a greater margin than previous models, making it more useful than its predecessors. This advance in PDR is primarily due to the incorporation of temporal PDR performance and the use of sidechains throughout the entire route estimation process. As a consequence of these performance enhancements, the proposed model is applicable to a wide range of IoV routing application scenarios, even in the face of large-scale attacks.

## 5. Conclusions & Future Scope

This paper concludes with a novel routing model designed specifically for Internet of Vehicles (IoV) deployments. To enhance the efficacy of routing in IoV scenarios, the proposed model integrates several features, including efficient fan-shaped clustering, trust-based routing, quality of service (QoS) considerations, security-aware side-chaining, and distance measurements.

The evaluation and analysis conducted in this paper demonstrate the efficacy of the proposed model in comparison to extant routing models, namely P2SF, LS D3S, and Se MAV, in terms of several important performance metrics. Using the proposed model results in significant improvements in latency, energy consumption, throughput, and packet delivery ratio (PDR).

In terms of delay, the proposed model obtains a reduction of 12.5% compared to P2SF, 14.5% compared to LS D3S, and 8.5% compared to Se MAV compared to the existing models. This latency reduction is especially advantageous for real-time scenarios and high-speed routing, allowing for quicker and more efficient communication in IoV deployments. It is determined that the incorporation of distance measurements into routing fitness functions is the primary contributor to this enhancement for different scenarios.

The proposed model outperforms the extant models in terms of energy consumption by achieving energy reductions of 15.5% versus P2SF, 19.5% versus LS D3S, and 18.5% versus Se MAV. This reduced energy consumption is advantageous for long-lasting network routing deployments, particularly in large networks. The decrease in energy consumption can be attributed to the utilisation of residual energy during the modelling of routing fitness functions.

In addition, the evaluation proves that the proposed model has superior throughput performance. It improves throughput by 15.2% relative to P2SF, 19.4% relative to LS D3S, and 24.5% relative to Se MAV. This increased throughput makes the proposed model an excellent candidate for high-speed routing deployments. Consideration of each route's temporal throughput performance contributes to this significant improvement.

In addition, the proposed model has a higher packet delivery ratio (PDR) than extant models. It attains a PDR that is 2.3% superior to P2SF, 3.5% superior to LS D3S, and 8.5% superior to Se MAV. This enhancement in PDR is advantageous for high-efficiency routing installations, thereby making the proposed model more practical and dependable than its predecessors. Principally responsible for this improvement is the incorporation of temporal PDR performance throughout the complete route estimation process.

In addition, the efficacy of the proposed model is evaluated under various attack scenarios, including Sybil, Distributed Denial of Service (DDoS), Finney, and Man-in-the-Middle attacks. The results indicate that the proposed model maintains its superiority over existing models even when under attack. In comparison to P2SF, LS D3S, and Se MAV, it has decreased latency, energy consumption, throughput, and PDR. The incorporation of side-chains and the use of highly efficient side-chains derived from an analysis of temporal miner performance contribute to the enhanced performance under attack scenarios.

The paper concludes with a thorough evaluation of the proposed routing model for IoV deployments. The results demonstrate its efficacy in reducing delay, energy consumption, improving throughput, and augmenting PDR in comparison to existing models. The inclusion of distance measurements, residual energy utilisation, temporal performance considerations, and security-aware side-chaining all contribute to the proposed model's superior performance. This research contributes to the development of routing protocols in IoV and offers valuable insights for the design of efficient and secure routing solutions for future IoV deployments.

**5.1. Future Scope**

This article contributes significantly to the field of routing in Internet of Vehicles (IoV) environments. Nonetheless, future research and development can enhance and expand the proposed model in a variety of ways. Here are some possible future applications of this paper:

While the proposed model exhibits enhanced performance in comparison to extant models, there is space for further performance optimisation. Future research can investigate sophisticated algorithms and techniques to improve the performance of fan-shaped clustering, trust-based routing, and side-chaining. This may entail the development of innovative routing metrics, adaptive clustering algorithms, and intelligent resource allocation strategies in order to optimise latency, energy consumption, throughput, and packet delivery ratio.

Scalability and Resilience: As IoV deployments continue to increase in scope and complexity, it is essential to ensure that the proposed model remains scalable and resilient. Future research can investigate techniques such as hierarchical clustering, distributed routing management, and load balancing mechanisms to address scalability issues. Exploring intrusion detection and prevention mechanisms, anomaly detection algorithms, and resilient routing protocols can further enhance a system's resilience against various types of attacks.

Integration of Artificial Intelligence (AI): Adding AI techniques to IoV routing can provide additional benefits. Future research could investigate the use of machine learning algorithms for intelligent decision-making, adaptive routing, and anomaly detection. Approaches based on artificial intelligence can enable the model to learn and adapt to shifting traffic patterns, network conditions, and security threats, thereby improving the routing system's overall performance and resilience.

Differentiation of Quality of Service (QoS): QoS differentiation is required in IoV deployments to support diverse applications with varying requirements. Future research can concentrate on the development of mechanisms that provide differentiated QoS guarantees based on application-specific requirements. This may involve the design of efficient traffic prioritisation schemes, bandwidth allocation strategies, and dynamic QoS provisioning mechanisms to ensure optimal performance for various IoV applications, including safety-critical communications, multimedia streaming, and real-time data exchange.

Standardisation and Interoperability: As IoV deployments involve a variety of vehicles, sensors, and infrastructure components from various manufacturers, it is essential to ensure interoperability and standardisation. Future research can investigate the creation of standardised protocols and interfaces that facilitate the integration and communication of heterogeneous IoV devices and networks. This can promote the widespread adoption and interoperability of the proposed routing model across all IoV deployments.

Real-world Deployments and Field Tests: To validate the efficacy of the proposed model in real-world scenarios, it is essential to conduct large-scale field tests and deployments. Future research may concentrate on implementing and evaluating the proposed model in actual IoV testbeds and experimental environments. This can provide valuable insight into the practical difficulties, performance trade-offs, and deployment considerations of the proposed model and its associated mechanisms.

Enhancements to Privacy and Security Privacy and security are crucial concerns for IoV deployments. Future research may investigate the incorporation of advanced cryptographic techniques, mechanisms for protecting privacy, and secure communication protocols into the proposed model. This can ensure the confidentiality, integrity, and authenticity of the exchanged data as well as safeguard the privacy of vehicle and user identities, thereby improving the routing system's overall security posture.

In conclusion, future research directions for this paper include further performance optimisation, ensuring scalability and robustness, integrating AI techniques, differentiating QoS, promoting standardisation and interoperability, conducting real-world implementations, and enhancing privacy and security aspects. These research directions will contribute to the ongoing development of efficient and secure Internet of Vehicles routing solutions.